\documentclass[preprint]{aastex}
\begin{document}
\title{Thermodynamics of a Higher Order Phase Transition: Scaling 
Exponents and Scaling Laws}
\author{P. Kumar$^\ast$ and A. Saxena$^\dag$}
\affil{$^\ast$Department of Physics, University of Florida,
Gainesville, FL 32611}
\affil{$^\dag$Theoretical Division, Los Alamos National Laboratory, 
Los Alamos, New Mexico 87545 }
\begin{abstract}
The well known scaling laws relating critical exponents in a second 
order phase transition have been generalized to the case of an arbitrarily higher order phase transition.  In a higher order transition, such as one suggested for the superconducting transition in Ba$_{0.6}$K$_{0.4}$BiO$_3$ and in Bi$_2$Sr$_2$CaCu$_2$O$_8$, 
there are singularities in higher order derivatives of the free energy.  A relation between exponents of different observables has been found, regardless of whether the exponents are classical (mean-field theory, no fluctuations, integer order of a transition) or not (fluctuation effects included).  We also comment on the phase transition in a thin film.   
\end{abstract}
\section{Introduction}
In a recent letter \citep{kumar,hall1}, there is a proposal that the 
superconducting phase transition in Ba$_{0.6}$K$_{0.4}$BiO$_3$ (BKBO)  
is of order four, in the sense defined by \citet{ehren}.  Thus the first three derivatives of the free energy with respect to the temperature $T$, the entropy $S$, the specific heat $C$ and the temperature derivative of the specific heat $\partial C/\partial T$ are continuous across the phase transition.  As are the first three derivatives with respect to the magnetic field $H$; the magnetization $M$, the susceptibility $\chi$ and $\partial\chi/\partial H$. The fourth order derivatives, namely $\partial^2C/\partial T^2$ and $\partial^2\chi/\partial H^2$ are discontinuous.  The discontinuity may even be a small power-law singularity such as a $\lambda$-point anomaly but in $\partial^2C/\partial T^2$ and $\partial^2 \chi/\partial H^2$. 

There are several materials \citep{kumar2} which are possible candidates for a higher order phase transition (see the discussion at the end of this paper).  One class consists of Barium and Bismuth based cubic perovskites, doped with K, Rb or Sr.  One of the earliest anomalies about the original material, Ba(Pb$_{1-x}$Bi$_x$)O$_3$ (BPBO) was an 
absence \citep{methf} of the usual specific heat discontinuity at 
the superconducting phase transition.  Many of the materials listed in 
\citep{kumar2} have not been studied for their specific heat yet, nor for any 
anomalous feature in the associated magnetization.  For a second order 
phase transition, the phase boundary in the $H$-$T$ phase diagram is 
described by, 
$$\left(\frac{\partial H}{\partial T}\right)^2 = \frac {\Delta C} 
{T_c\Delta\chi},  \eqno(1)$$
where $T_c$ is the transition temperature.  If $\Delta C$ is zero then 
either the phase boundary is flat, or $\Delta\chi$ must also be zero 
and the transition is of a higher order.  In most materials which are 
paramagnetic\footnote{In the Abrikosov state, the magnetization is negative but the susceptibility is paramagnetic and is usually larger than the normal state paramagnetism value.  Thus one often finds a discontinuity in $\chi$ at $H_{c2}$. In contrast, the materials discussed here are often diamagnetic.  It is easy then to smoothly connect the magnetization across the superconducting transition so that around the upper critical field $H_{c2}$, the susceptibility is zero on both sides of the transition.  The discontinuity is then in the nonlinear susceptibility as expected for a higher order transition.}, it is quite unlikely that $\Delta\chi = 0$. 
Most of the materials discussed above however are diamagnetic in the normal state.  Here the condition $\Delta\chi = 0$ may be more easily satisfied, thus making them possible candidate materials for a higher order phase transition.

Following an extensive analysis of the temperature dependence of the specific heat near the superconducting transition, \citet{junod} have proposed that the superconducting transition in Bi2212 (Bi$_2$Sr$_2$CaCu$_2$O$_8$) is similar to the transition one encounters in Bose-Einstein condensation in an ideal Bose gas. From the specific perspective here, this is an example of a third order phase transition.  We elaborate on this topic below in the discussion section.    

The purpose of this paper is to further explore the thermodynamics 
of a $p^{th}$ order phase transition.  Let us define the terms: the condensed state free energy $-F(T)\sim t^{p-\mu}$, where ${t=(1-T/T_c)}$, $p$ is an integer and $\mu\ne0$, a small correction to $p$.  In general, $\mu$ should also have a subscript $p$ signifying its dependence on $p$.  We will often omit this subscript for convenience. The term classical will refer to the case $\mu=0$ and the term non-classical will include effects of a non-zero $\mu$.  For example, for a second order transition $p=2$ and $\mu = \alpha$, where $\alpha$ is the specific heat exponent. The term non-classical then encompasses all fluctuation effects related to the ground state.  In the results reported in refs. 1 and 2, the exponent for the temperature dependence 
of the free energy is 3.6.  The suggestion thus being that $p=4$ and 
$\mu = -0.4$,  the specific heat exponent $\alpha = -1.6 < 0$, i.e., there is no divergence in the specific heat but that the second derivative of the specific heat $\partial^2C/\partial T^2$ has a $\lambda$-like divergence with an exponent $\mu$.  Similarly, there may be a $\lambda$-like divergence in $\partial^2\chi/\partial H^2$ with a corresponding exponent.

We have focused the discussion here to a free energy depending on 
magnetic field and temperature.  In general, there are other mechanical 
variables which can be accommodated by simply replacing the thermodynamically conjugate pair ($M$, $H$) by the appropriate combination (for example volume $V$ and pressure $P$).  Moreover, we will also discuss two features which are specific to a superconductor.  In a superconductor, the magnetic field influences the charge motion via a gauge coupling.  Indeed, if the Cooper pairs are assumed to be in a spin singlet state, as we assume in the following, then the magnetism is mostly due to the orbital contribution and the susceptibility has a special temperature dependence.  The scaling exponents are then specific to that case.  The second superconductivity feature here is the existence of two different critical fields.  In the $H$-$T$ plane, flux expulsion occurs at $H_{c1}(T)$, which is lower than the upper critical field $H_{c2}(T)$ where superconductivity is destroyed.  Both of these critical fields vanish at $T_c$.  In a second order phase transition they are both linear in reduced temperature and their ratio, the Landau parameter $\kappa$, is a constant, independent of temperature.  In a higher order transition \citep{kumar} $\kappa$ is temperature dependent, diverging at $T_c$.  The lower critical field also is a measure of the superfluid density through the London penetration depth.       

Thus the well-known scaling laws appropriate for a second order phase 
transition can be generalized for an arbitrary order phase transition as described in Sec. II below.  In Sec. III, we study a relationship between the specific heat and the London penetration depth, which is a relation between the exponents $p-\mu$ and an exponent for the temperature dependence of the lower critical field $H_{c1}(T)$.  Section IV contains a discussion of finite size effects.  In this section we also discuss a Kosterlitz-Thouless type phase transition, the binding-unbinding of a vortex-antivortex pair, as well as the irreversibility effects associated with the melting of a flux lattice.  Finally, Sec. V contains a summary of our results including a discussion of possible candidate materials and field theory models for a higher order phase transition.

\section {Scaling Laws}

Scaling laws are consistency checks based on the (magnetic) 
field-temperature dependence of the experimental observables.  The 
free energy is considered as a function of temperature $T$ and a 
mechanical variable, say, the magnetic field $H$.  The exponents are 
defined as follows:
$$ C=-T\frac{\partial^2 F}{\partial T^2} = t^{-\alpha}; 
~~m = -\frac{\partial F}{\partial H}=t^{\beta}; 
~~\chi = \frac{\partial m}{\partial H} = t^{-\gamma};
~~m(T=T_c) = H^{1/\delta}, \eqno (2)$$
where $t=(1-T/T_c)$ is the reduced temperature, $C$ is the specific heat and $m$ denotes the magnetization.  These exponents\footnote{We know of only one book \citep{pippard} which considers a higher order transition.} are related \citep{baker} via, inter alia, the Rushbrooke inequality \citep{rushbrooke} 
$$ \alpha + 2\beta + \gamma \geq 2. \eqno (3)$$
This inequality was originally derived for a second order phase 
transition.  However, it is valid for transitions of all orders.  
For a higher order transition, the singular derivative is the $p$th 
derivative of the free energy with respect to temperature or magnetic 
field.  By definition, the exponent of the thermal derivative is $\mu$.  
By construction, $p-\mu = 2 - \alpha$ and Rushbrooke inequality becomes 
$\mu + 2\beta + \gamma \ge p$.  But this is really not yet the 
relationship we are seeking, i.e., the one between the exponents of 
possibly singular observables.  In order to focus on that relationship 
it is easier to study the equality. 

Thus in a $p$th order phase transition, the singular 
derivative with respect to magnetic field (as an example of a 
mechanical variable) is also the $p$th one, $N_p=\partial^pF/ 
\partial H^p=\partial^{p-2}\chi/\partial H^{p-2}$.  Let us 
consider therefore (while defining $\zeta_p$ and $\kappa_p$; the exponent 
$\kappa_p$ need not be confused with the Landau parameter $\kappa$ above) 
$$F=-t^{p-\mu}\simeq-(H_{c_2}-H)^{p-\zeta_p}, 
~~~H_{c_2}=t^{x_2} , \eqno(4) $$ 
$$ N_p=\frac{\partial^p F}{\partial H^p} 
=t^{-\kappa_p} . \eqno(5) $$ 
It follows therefore that $\kappa_p = x_2\zeta_p$.  By identifying 
the most singular terms, we can derive the following identities: 
$$p-\mu=x_2(p-\zeta_p),~~\beta=x_2(p-\zeta_p -1),~~
{\delta}^{-1}=p-\zeta_p-1,~~\gamma=-x_2(p-\zeta_p-2). \eqno(6) $$

These can be further processed to produce the following identities:
$$\beta+\gamma=x_2, ~~\zeta_p=p-\frac{(\gamma+2\beta)}{(\gamma+\beta)},
~~ \kappa_p=(p-1)\gamma+(p-2)\beta. \eqno (7) $$
Here $x_2$, the exponent for the upper critical field $H_{c2}(T)$ is 
an observable, as are $\beta$ and $\gamma$.  The latter two, in general, 
correspond to non-singular (if $p>2$) observables.  The first expression 
in Eq. (7) is a scaling law relating temperature dependence of exponents 
between two different experiments.  Similarly the last expression too 
relates different experiments.  The middle expression is an important 
algebraic link between the two.

Finally, by eliminating $x_2$ we can derive the scaling relations: 
$$(p-1)\mu_p+p\beta_p+\kappa_p=p(p-1) , \eqno(8) $$ 
$$ \kappa_p=\beta_p[(p-1)\delta_p-1] , \eqno(9) $$ 
$$ \mu_p=p-\beta_p(\delta_p+1) , \eqno(10) $$ 
$$ (\delta_p+1)\kappa_p+[(p-1)\delta_p-1](\mu_p-p) = 0 . \eqno(11) $$ 
Recall that the classical values here are $\mu_p=\kappa_p=0$; 
$\beta_p=1/\delta_p=p-1$ and $\gamma_p=2-p$.  These might look unfamiliar to readers more accustomed to critical exponents in magnetism where the coupling to magnetic field appears through Zeeman interaction ($-M.H$).  In the case of a gauge coupling, appropriate for a superconductor, the classical exponents are exactly as described above for $p=2$.  Equation (8) is the new version of the Rushbrooke  \citep{baker,rushbrooke} equality while Eq. (9) is a generalization of ``Widom's law" \citep{widom}.  Finally Eq. (10) is the well known Griffith's law\citep{griffiths}. Thus, if the exponents are non-classical ($\mu\ne0$) they are all interconnected via the scaling laws. 

We have not discussed the exponents $\nu$ and $\eta$ and the associated analog of the Buckingham-Gunton inequality \citep{baker,buck}.  These exponents involve spatially varying features and remain subjects for future discussion.  However, a calculation of non-classical exponents, in principle, is no more difficult than it has been for $p=2$. Put differently, it is not advisable to consider a free energy with a power law temperature dependent exponent of 3.6 as $2+1.6$, which would describe a second order phase transition with a specific heat exponent $\alpha = -1.6$, due to fluctuations.  Indeed much of the machinery for a calculation of exponents is designed with small $\alpha$ in mind.   More likely, the right course of action is to consider a different mean field theory followed by a consideration of fluctuations about that theory, i.e., the free energy exponent should be viewed as $4-0.4$, with the mean field corresponding to the higher integer.    

\section {Other Thermodynamic Identities}

The above relations were originally derived to relate exponents which 
incorporated fluctuation effects around a second order transition.  In 
the following we consider other identities which are characteristic of 
a mean field description for a second order transition.  These can   
not clearly be extended to include fluctuations but extension to higher 
order transitions is possible.  For example near $T_c$, it is possible 
to derive a relationship between the specific heat (in its temperature 
dependence) and the London penetration depth $\lambda(T)$.  That a 
result such as this should exist is clear from the way the magnetic 
field couples to the system.  Consider for example the free energy for 
a second order phase transition in the presence of a magnetic field, 
$$ F=-a_0t|\psi|^2+b|\psi|^4+c\left|\left(\nabla+\frac{2\pi i} 
{\phi_0}{\bf A}\right)\psi\right|^2+\frac{1}{2\mu_0}(\nabla\times 
{\bf A})^2 . \eqno(12) $$ 
Here $t=1-T/T_c$; $a_0$, $b$ and $c$ are all positive constants.  
The complex order parameter $\psi$ is the independent variable which 
takes its ground state value $\psi_0$ by minimizing the free energy 
$F$.  $\bf A$ is the vector potential so that the magnetic induction 
{\bf B} is given by ${\bf B}=\nabla\times{\bf A}$.  Finally, $\phi_0= 
h/2e$ is the superconductor flux quantum and $\mu_0$ is the magnetic 
permeability.  If we minimize Eq. (12) 
with respect to the vector potential to derive the Euler-Lagrange 
equation for ${\bf A}({\bf r})$, we get 
$$\frac{1}{\lambda^2}=2\mu_0c\left(\frac{2\pi}{\phi_0}\right)^2 
|\psi_0|^2 . \eqno(13) $$ 
Thus, near $T_c$, all temperature dependence in $\lambda^{-2}$ comes 
from the temperature dependence of the order parameter $|\psi|^2$. 
The coupling between the vector potential {\bf A} and the order 
parameter is the gauge coupling and it is clear that this is the 
only temperature dependence possible for $\lambda^{-2}$ (near 
$T_c$, the quasiparticle contribution to $\lambda^{-2}$ is 
expected to be small compared to the order parameter contribution 
here).  The free energy $F(T)$ can be written as 
$$F(T)=-b\psi_0^4; ~~~~\psi_0^2=\frac{a_0t}{2b},~~~t>0 
~~(=0, ~~~t<0) . \eqno(14)  $$ 
Thus, by expressing $\psi_0$ in terms of $\lambda$, a simple result 
emerges, 
$$C(T)=-T\frac{\partial^2F}{\partial T^2}=\frac{bT}{(2\mu_0c)^2} 
\left(\frac{\phi_0}{2\pi}\right)^4\frac{\partial^2}{\partial T^2} 
\lambda^{-4} . \eqno(15) $$ 
Apart from the constant $b$ which appears in $C(T)$ and $c$, which 
is a measure of the gradient coupling, the other constants such 
as $\phi_0$ and $\mu_0$ are fundamental. 

Equation (15), to our knowledge has not been widely used. 
It is specific to the form of the gauge coupling and it is easy to 
generalize to a $p$th order phase transition.  The free 
energy is believed to be:  
$$F(T)=-a_0t|\psi|^{2(p-1)}+b|\psi|^{2p}+c\left|\left(\nabla+\frac 
{2\pi i}{\phi_0}A\right)\psi^{(p-1)}\right|^2 , \eqno(16) $$ 
where we have omitted the magnetic field energy term, the last term 
in Eq. (12), for brevity.  Here the singular derivative of the free 
energy is $M_p=\partial^pF/\partial T^p=\partial^{p-2}C/\partial T^{p-2}$, 
rather than the specific heat itself.  Thus [$l=p/(p-1)$]; 
$$M_p=\frac{b}{(p-1)(2\mu_0c)^l}\left(\frac{\phi_0}{2\pi} 
\right)^{2l}\frac{\partial^p}{\partial T^p}\lambda^{-2l} . 
\eqno(17) $$ 
Note that for $p=2$ Eq. (17) reduces to Eq. (15). 

It is tempting to wonder whether Eq. (17) provides a scaling 
relation between the exponent for free energy (or specific 
heat) and one for the London penetration depth $\lambda(T)$ 
$(\lambda^{-2}\sim t^{x_1})$. Our derivation so far is limited to 
the mean field regime.  
Let us define scaling exponents so that near $T_c$ 
$$F(T)\sim-t^{p-\mu}, ~~~H_{c_1}=t^{x_1}, ~~~H_{c_2}\simeq 
t^{x_2} , \eqno(18) $$ 
then we expect from $H_c^2=H_{c_1}H_{c_2}$, $x_1+x_2=p-\mu$ 
while from Eq. (17), we have $x_1l=p-\mu$ so that 
$$ x_1=\frac{(p-\mu)}{p} (p-1), ~~~ x_2=\frac{p-\mu}{p} . 
\eqno(19) $$ 
For $p=2$, $x_1=x_2=1-\mu/2$.  For $p=4$, the classical values 
are $x_1=3$, $x_2=1$. 

We have measurements \citep{kumar,hall1} for both $x_1$ and $x_2$: 
$x_1=3.03\pm0.16$ and $x_2=1.21\pm0.02$.  However, there are 
problems.  From Eq. (19); we see that, in order to be consistent 
with the temperature dependence of $H_{c2}$, $x_2<1$.  Another 
manifestation would be $x_1+x_2<4$; but the measured values 
do not satisfy that. This is nearly but not quite satisfied by the 
experimentally measured exponents.

This may well be due to the fluctuation effects.  The London penetration depth is a measure of the superfluid density.  But the temperature dependence of the superfluid density is better defined via the penetration depth.  When we replace the order parameter in Eq. (14) 
in favor of $\lambda$ by using Eq. (13), we have restricted the 
identity to mean field theory.  Moreover, $x_1$ is determined from 
the temperature dependence of the lower critical field $H_{c1}(T)$ 
which contains logarithmic corrections to the temperature dependence 
of $\lambda$.  In view of the divergent temperature dependence 
\citep{kumar,kumar2} of the Landau parameter $\kappa = \lambda/\xi$, the 
exponent for 
$\lambda$ may be smaller than $x_1$; $\xi$ denotes the coherence 
length.  We have here a relationship between the specific heat $C(T)$ 
and the London penetration depth $\lambda(T)$.  An objective of the  
discussion here is to motivate precise and direct measurments of 
$\lambda(T)$

\section{Transition in a Film and Melting of vortices}

A feature of the Ginzburg-Landau model for a $p$th order 
transition described by Eq. (16) is the temperature dependence 
of the superfluid density $\rho_s\propto|\psi|^{2(p-1)}\propto 
t^{p-1}$.  The superfluid density may be identified by the 
kinetic energy being $\frac{1}{2}\rho_sv_s^2$ where $v_s=\frac 
{\hbar}{m^\ast}(\nabla\phi)$, $\phi$ denotes the phase of the order 
parameter and $m^\ast$ is the effective mass of the Cooper pair.  One 
consequence is the nature of the Kosterlitz-Thouless transition 
\citep{tinkham} in a thin film with thickness $d$.  In two dimensions, a vortex-antivortex pair can unbind above a temperature $T_{KT}$, leading to s destruction of the superconducting state. The expression for the transition temperature $T_{KT}$ is given as 
$$kT_{KT}=\frac{\phi_0^2}{32\pi^2}\frac{d}{\lambda^2(T_{KT})} $$ 
or 
$$\frac{T_{KT}}{T_c}=\frac{d}{d_k}\left(1-\frac{T_{KT}}{T_c}\right) 
^{p-1}\simeq\frac{d}{d_k}\left(1-(p-1)\frac{d}{d_k}+...\right) , 
\eqno(20) $$ 
where $d_k$ is a scaled length.  There are two major consequences.  
$T_{KT}$ in general is smaller for a higher order phase transition 
than for a second order one.  But the ratio $\Delta\rho_s/T_{KT}$, a universal 
constant, remains the same constant here.  The decrease in $\Delta\rho_s$
is entirely due to the decrease in $T_{KT}$.

In three dimensions, the analog of K-T transition is flux-melting,
\citep{tinkham,houghton} particularly in anisotropic, high $T_c$ 
superconductors.  In the $H$-$T$ phase space, the melting of vortices 
occurs at a phase boundary which has been calculated by among others
\citet{Nelson} and \citet{houghton}.  The qualitative results can be encapsulated in a simple expression \citep{tinkham} for the flux melting line.
$$B_m (T) \simeq {\lambda}^{-4}. \eqno(21)$$
Given the strong temperature dependence of $\lambda$, this field 
cannot be identified with the observed irreversibility line.  We will 
defer this subject to a later exploration. 

\section{Discussion}

The principal result we have here is the derivation of scaling laws, 
Eqs. (8)-(11), appropriate for the exponents in a higher order phase 
transition.  In addition, we have explored the possibility of other 
identities which at the moment seem to be restricted to a mean field 
description but may lend themselves to a more general analysis.

This is in fact related to a more general issue in superconductivity.  In the conventional superconductors, the transition (to the extent known) was second order, and a mean field theory was quite sufficient, the exponents were all classical.  Then came the high $T_c$ superconductors and dominance of fluctuations.  Thus, in high $T_c$ superconductors, it is often argued \citep{blatter} that there is no $H_{c2}(T)$.  The coherence length is very small and the fluctuations are dominant.  The transition is determined by whether the vortices are pinned in which case the resistivity is zero, or not in which case the system becomes a normal metal.  This is qualitatively different physics where macroscopic defects play a critical role.  Scaling laws, developed to describe moderate effects of fluctuations, have apparently no relevance in this case. 

But in BKBO the coherence length, as determined by $H_{c2}(T)$ is of 
order 6 nm.  This is not small.  That the transition is higher order 
does not necessarily mean that fluctuations have altered the physical 
landscape.  On the contrary, based on the currently available information,  the fluctuations here may well be moderate and scaling laws will likely be useful.

Doubts have been raised with regards to whether the transition in BKBO is of higher order.  For example \citet{phillips} have reported a small discontinuity in the specific heat at the superconducting $T_c$.  If there is a discontinuity in the specific heat then there is no need to invoke a higher order phase transition.  But the order of the transition \citep{kumar} is determined by the temperature dependence of the thermodynamic critical field. The Berkeley results \citep{phillips} also indicate that the discontinuity disappears at magnetic fields of order 3 Tesla.  This is much smaller than any other measurement of upper critical field (see other citations in \citet{hall1}).  The case developed in \citet{kumar} is internally consistent and is based on 
measurements of field and temperature dependent magnetization.  

Moreover, in a wide ranging study of a number of high $T_c$ materials, 
 \citet{junod} suggest that the specific heat anomaly at the 
superconducting transition in Bi2212 compounds is distinctly different 
from the mean field behavior in conventional superconductors as well 
as a $\lambda$-point anomaly in Y123 (YBa$_2$Cu$_3$O$_7$).  In particular, they suggest that the specific heat in Bi2212 appears similar to that near a Bose-Einstein condensation (BEC).  In as much BEC (in an ideal Bose gas) may be seen as a third order phase transition\footnote{The compressibility of a neutral Bose-Einstein condensate is infinite. Thus \citet{london} has argued that the BEC in an ideal Bose gas in the $P$-$T$ plane is a first order phase transition.  But in a non-ideal system, we might not have this anomalous behavior.} , Bi2212 should be considered 
as a candidate material as well.  As \citet{junod} note, the 
specific heat in Bi2212 is continuous.  It is the temperature derivative of the specific heat which is singular at $T_c$ with an exponent of $\mu_3 = 0.33$.   

Finally, we note that a third order phase transition has been proposed 
in the large-$N$ limit of the two-dimensional U($N$) lattice gauge 
theory with variation in the coupling constant (or analogously the  
temperature in statistical mechanics) \citep{gross}.  A discussion of 
the third order transition in an associated chiral model and pertinent 
exponents is given in \citet{camp}.  A possible third order phase 
transition in Invar type alloys \citep{shiga} and a fourth order 
transition in the antiferromagnetic Blume-Capel model \citep{wang} have 
been reported in the literature.  However, no attempt was made to 
either provide a free energy or derive the scaling exponents.  
 
\section{Acknowledgment}
This work was supported in part by the U.S. Department of Energy and 
in part by the National Science Foundation.  We are grateful to R. 
Goodrich, D. Hall, D. Hess, A. J. Houghton, G. Stewart and J. D. Thompson for discussions.


\end{document}